# Controlled micro/nano-dome formation in proton-irradiated bulk transition-metal dichalcogenides

Davide Tedeschi,[1] Elena Blundo,[1] Marco Felici,[1] Giorgio Pettinari,[2] Boqing Liu,[3] Tanju Yildrim[4], Elisa Petroni,[1] Chris Zhang,[4] Yi Zhu,[4] Simona Sennato,[5] Yuerui Lu,[4*] Antonio Polimeni[1*]

[1] Dipartimento di Fisica, Sapienza Università di Roma, 00185 Roma, Italy.

[2] Institute for Photonics and Nanotechnologies, National Research Council, 00156 Roma, Italy.

[3] Research School of Engineering, College of Engineering and Computer Science, The Australian National University, Canberra, ACT2601, Australia.

[4] College of Chemistry and Environmental Engineering, Shenzhen University, P. R. China

[5] Institute for Complex Systems, National Research Council, 00185 Roma, Italy.

[*] Correspondence to: yuerui.lu@anu.edu.au; antonio.polimeni@roma1.infn.it

**At the few-atom-thick limit, transition-metal dichalcogenides (TMDs) exhibit strongly interconnected structural and optoelectronic properties. The possibility to tailor the latter by controlling the former is guaranteed to have a great impact on applied and fundamental research. As shown here, proton irradiation deeply affects the surface morphology of bulk TMD crystals. Protons penetrate the top layer, resulting in the TMD-catalyzed production and progressive accumulation of molecular hydrogen in the first interlayer region. This leads to the blistering of one-monolayer thick domes, which stud the crystal surface and**

**locally turn the dark bulk material into an efficient light emitter. The domes are stable (>2-year lifetime) and robust, and host strong, complex strain fields. Lithographic techniques allow to engineer the formation process so that the domes can be produced with well-ordered positions and sizes tunable from the nanometer to the micrometer scale, with important prospects for so far unattainable applications.**

## 1. Introduction

The notable transformation of the electronic properties of transition metal dichalcogenides (TMDs) when reduced to a single X-M-X plane (X: chalcogen; M: metal)[1] makes them suitable for flexible, innovative optoelectronic devices[2,3,4] and transistors[5]. Like graphene, few-layer TMDs can also withstand surprisingly large mechanical deformations[6,7,8,9], which, coupled to the material's electronic structure, would enable the observation of non-dissipative topological transport, provided a periodic modulation of strain is attained[10,11,12,13]. Finally, TMD monolayers (MLs) and nanostructures are also important for their catalytic role in the cost-effective production of hydrogen[14,15,16]. These examples share the need to achieve spatial control of the material's properties, over sample regions with size ranging from the nano[14,16] to the micrometer[16] scale lengths.

In this study, we present a route toward the patterning of TMDs based on the effects of low-energy proton irradiation[17] on the structural and electronic properties of *bulk* $WS_2$, $WSe_2$, $WTe_2$, $MoS_2$, $MoSe_2$ and $MoTe_2$. Suitable irradiation conditions trigger the production and accumulation of $H_2$ just beneath the first X-M-X basal plane, leading to the localized exfoliation of the topmost monolayer and to the formation of spherically shaped domes. Structural and optical characterizations confirm that these domes are typically one ML-thick and contain $H_2$ at pressures in the 10-100 atm range, depending on their size. Such high pressures induce strong and complex strain fields acting on the curved X-M-X planes, which are evaluated by means of a mechanical model. The domes' morphological characteristics can be tuned by lithographically controlling the area of the sample basal plane participating in the hydrogen production process. This results in the unprecedented fabrication of robust domes with controlled position/density and sizes tunable from the nanometer to the micrometer scale, that, by virtue of their inherently strained nature and geometry, might prompt a variety of applications.

## 2. Creation of light-emitting micro-/nano-domes

### 2.1 Proton-irradiation of bulk TMDs

The samples, consisting of thick (tens to hundreds of MLs) TMD flakes, were obtained by mechanical exfoliation, deposited on Si substrates, and afterwards proton-irradiated using a Kaufman source (see Methods). The top-right inset of Fig. 1a displays the optical microscope image of a large bulk $WS_2$ sample after irradiation with an impinging dose $d_H = 8\times10^{16}$ protons/cm$^2$. Unexpectedly for indirect-gap bulk $WS_2$, the sample exhibits strong photoluminescence (PL, see Methods) in the red wavelength region ($\lambda$~690 nm), as displayed in the main part of Fig. 1a and Supplementary Fig. S1. The bottom-left inset shows that the luminescence originates from circular spots with diameters varying between less than one (which is an upper limit due to the finite resolution of our optical setup; see Methods) to few μm. As evidenced by the atomic force microscopy (AFM, see Methods) image in Fig. 1b, these spots signal the presence of dome-shaped features, protruding from the irradiated crystal surface, with virtually perfect spherical shape. The average footprint radius of the features displayed in Fig. 1b is $R$=(715±60) nm—with maximal height $h_m$=(230±20) nm—but the dome size can be controlled by the irradiated proton dose: with $d_H=1.0\times10^{16}$ protons/cm$^2$ we obtained nanometer-sized structures, with average $R$=(82±20) nm and $h_m$=(25.6±5.6) nm (Fig. 1c). Fig. 1d shows an AFM image encompassing three domes formed on a $WS_2$ sample analogous to that shown in panel c. The corresponding room temperature micro-PL (μ-PL, see Methods) intensity map, detected at $\lambda_{det}$=689 nm and displayed in panel e, demonstrates the perfect match between the domes and the light-emitting spots. The μ-PL spectrum of one dome (singled out from the ensemble of panel a) is provided in Fig. 1f. The luminescence of an untreated $WS_2$ ML measured under the same excitation/collection conditions is shown for comparison. The PL intensity of the domes is typically larger than that of the exfoliated $WS_2$ MLs, indeed suggesting a one-ML thickness for the domes. This hypothesis is confirmed by measuring the μ-PL spectrum and the thickness of the outer layer of *exploded* domes, which are both fully consistent with that of the monolayer (as exampled for a $MoSe_2$ dome in Fig. S2). This is further supported by the strong *circular dichroism* measured for the PL emission of the domes (in a $WS_2$ dome the degree of circular polarization is >50% at 130 K, see Fig. S3A), which is a property stemming directly from the hexagonal symmetry of the first Brillouin zone of *monolayered* TMDs (see Fig. S3B)[18] and by second harmonic generation measurements, as discussed in the following. Based on the percentage of visible domes that double as strong light emitters (see Fig. S1), at least 85-90% of the domes can be estimated to be one-ML thick.

The energy of the PL peak (corresponding to the free-exciton recombination) of a typical (single-layer) $WS_2$ dome is 200 meV lower than that of the ML (see Fig. 1f), mainly because of the strain exerting on the dome's surface[7,8].

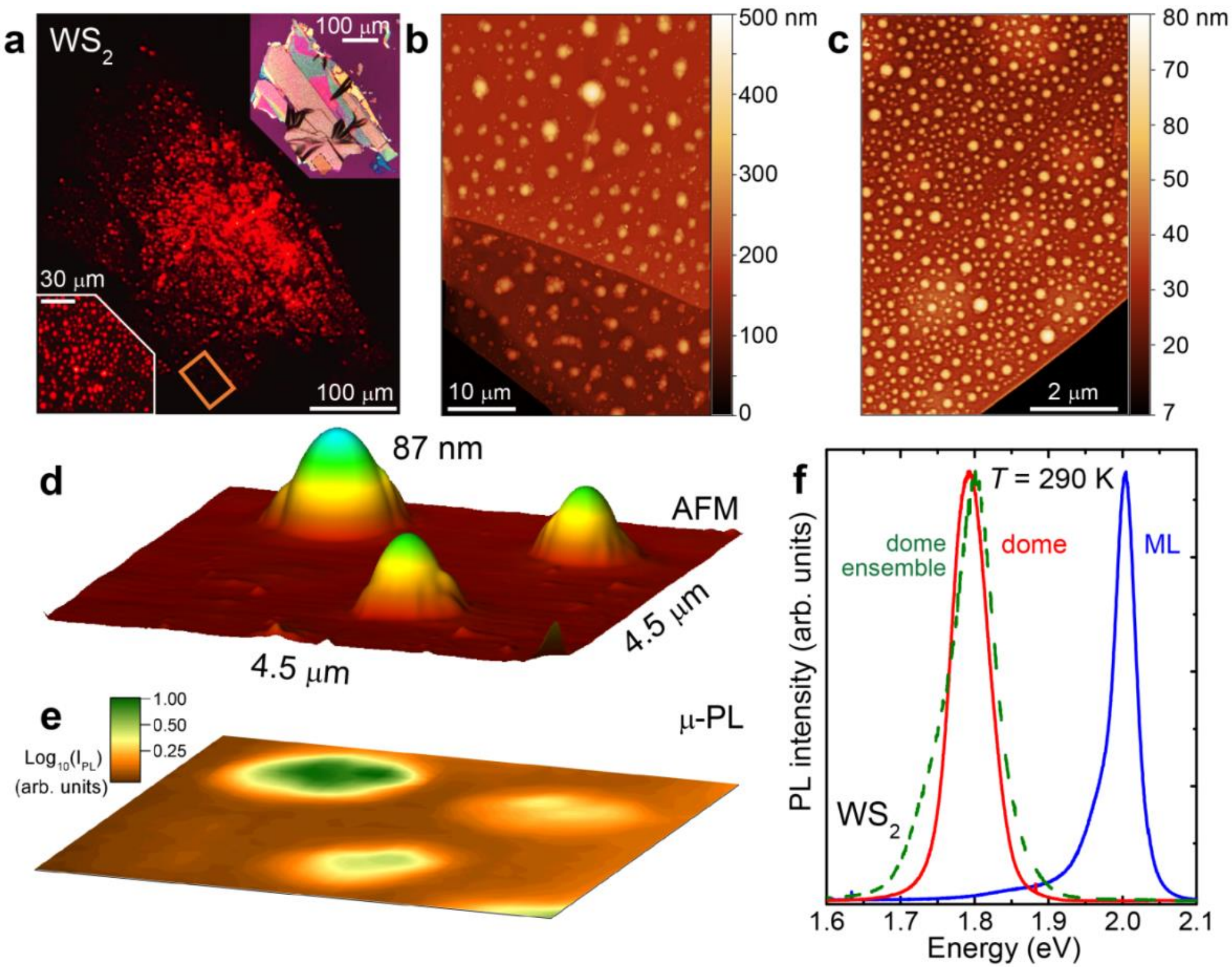

**Figure 1. Creation of Light Emitting Domes by proton irradiation**. **a,** Optical microscope image showing the laser-excited red luminescence of bulk $WS_2$ after irradiation with proton dose $d_H$=8×10$^{16}$ protons/cm$^2$. The top-right inset shows the same flake in absence of laser excitation. The bottom-left inset is a zoomed-in image of the main picture showing the round shape of the emitting spots. **b,** AFM image of the rectangular region highlighted in panel (**a**). Round-shaped features on the sample surface form after $H^+$ irradiation. **c,** AFM image of a bulk $WS_2$ flake $H^+$-irradiated with a dose 8 times smaller than in panels (**a**) and (**b**). **d,** AFM image of a limited region of a sample treated like that shown in (**c**) but on a different flake, where a smaller density of domes was fortuitously found. The maximal height reached by the domes is 87 nm. **e,** μ-PL mapping (detection wavelength equal to 689 nm) of the same region displayed in (**d**). The base-10 logarithm of the μ-PL intensity is shown as a false color scale (see color bar). **f,** μ-PL spectrum of a dome (red line) singled out from the ensemble displayed in (**a**); the blue line is the μ-PL spectrum of a $WS_2$ monolayer, whereas the green dashed line is the macro-PL spectrum of an ensemble formed by ~2500 domes. The spectra are peak-normalized for ease of comparison.

Notably, no significant linewidth broadening is observed when many such domes are measured all together, as demonstrated by the PL spectrum of a dome ensemble, displayed in Fig. 1f as a dashed line.

The spectral homogeneity of the light emitted by the domes suggests that the same average strain is present in each dome's "peel", irrespective of the dome size. The evidence shown in Fig. 1 demonstrates that indirect-gap *bulk* $WS_2$ can be turned into an efficient light emitter with no size restrictions, like those typically affecting exfoliated flakes or samples grown by chemical vapor deposition. We point out that the reported phenomenon is substantially independent of the specific $MX_2$ composition (see measurements on $H^+$-treated $MoSe_2$, $MoS_2$, $MoTe_2$, $WSe_2$ and $WTe_2$ in Figs. S4-S5) and is exclusively induced by the interaction of the material with protons. No effect was found in samples exposed to molecular hydrogen or ionized helium atoms under the same temperature and gas flow conditions.

### 2.2 Dome formation mechanism as a catalysis-driven process

Clues on the internal make-up of the domes can be derived by following their low-temperature evolution. Figs. 2a and 2b show, respectively, the 300 K and 4 K optical microscope images of a bulk $WS_2$ sample irradiated with $d_H$=8×10$^{16}$ protons/cm$^2$. At 300 K, many domes—featuring *iridescence*, as accounted for below—are visible. For $T \lesssim 30$ K the domes disappear, and the sample surface looks conspicuously flat. As $T$ is increased, at ~30 K, the domes bulge *suddenly* in ~10 mK (see insets of Fig. 2c and Movie 1). As illustrated in Fig. 2c, the transition temperature distribution was sampled over more than 500 domes obtained on several proton-irradiated flakes deposited on different substrates (see Methods). The resulting average transition temperature is (32.2±2.4) K, a number close to the critical temperature of $H_2$ (33.18 K) and thus compatible exclusively with the presence of molecular hydrogen inside the domes (see Supplementary Note SN1): when $T$ is sufficiently low/high, $H_2$ liquefies/boils, and the domes deflate/inflate always in the same position.

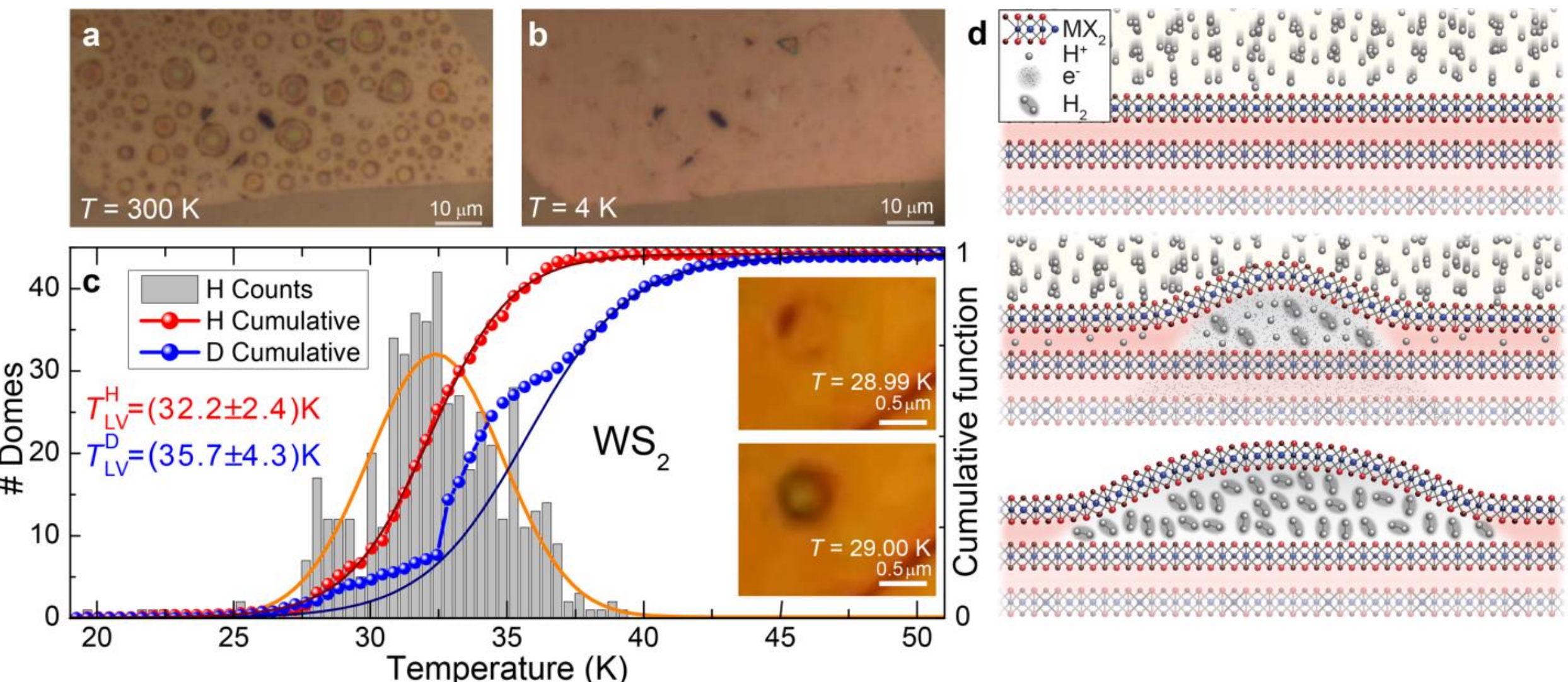

**Figure 2. Dome inflating/deflating process. a,** Optical microscope image of an $H^+$-irradiated ($d_H$=8×10$^{16}$ protons/cm$^2$) $WS_2$ sample at $T$=300 K. **b,** Same as (**a**) at $T$=4 K. **c,** Histogram of the transition temperature at which domes appear (left axis). The orange line is a Gaussian fit to the data. The red point-line is the cumulative function of the histogram (right axis). The solid wine curve superimposed is a fit to the cumulative function providing the average transition temperature $T^{H}_{LV}$ of proton-irradiated flakes. The blue point-line is the cumulative function of the histogram derived from deuteron-irradiated $WS_2$ flakes (see Supplementary Note SN1). The corresponding fit providing $T^{D}_{LV}$ is performed only on the part of the cumulative function in which H effects are negligible (see SN1). The insets are two optical microscope pictures of the same dome, recorded immediately before (top) and after (bottom) the transition temperature. **d,** Sketch of the process leading to the formation of domes caused by the local blistering of atomically thin material membranes.

The fluctuations observed in the liquid-vapor transition temperature ($T^{H}_{LV}$) are chiefly due to the spread in the size (and hence in the internal pressure) of the domes. To reinforce our hypothesis, we considered the study of isotopic effects in our system and we repeated our analysis on 500 more $WS_2$ domes, obtained by *deuteron irradiation*. The ($p$-$T$) phase diagrams of hydrogen and deuterium[19] show an isotopic shift in the temperature at which the liquid-vapor phase transition occurs of about 2.9 K (the exact value depending on the pressure of the gas). The details of our measurements are discussed in Supplementary Note SN1. The cumulative function of deuteron-irradiated samples id displayed in Fig. 2b (see Fig. SN1A for the histogram): differently from proton-irradiated samples, the behavior is in this case characterized by a composite shape, corresponding to the three different steps that are apparent in the cumulative function. We ascribe empirically these three steps to the presence of $H_2$, HD and $D_2$ molecules[20] within the domes. There are several possible explanations why H may be present also in deuterium-treated samples: 1) even the highest purity deuterium bottles contain a 0.25% of $H_2$ and HD molecules; 2) it has been reported that deuterons permeate through graphene and BN monolayers much slower than protons. At room temperature the areal conductivity of deuterons results to be 1/10 of that of protons[21]. This effect could be even much stronger in the case of TMDs, whose single layer is constituted by three atomic planes; 3) the hydrogen evolution reaction could be remarkably faster with respect to the deuterium evolution reaction because of the high mass difference between the two isotopes[22,23]. For deuteron-irradiated samples, a sigmoidal fit was thus performed only to the high-temperature data of the cumulative function shown, where no effect from hydrogen is expected. The fit estimates a transition temperature of $T^{D}_{LV} = (35.7 \pm 4.3)$ K, resulting in a difference between the centers of the distributions equal to (3.5±0.7) K, consistent with the (*p*-*T*) diagram of the two isotopes. Indeed, the measured transition temperatures are consistent only with the presence of $H_2$ (or its isotopes) within the domes (see Fig. SN1B). We can therefore hypothesize that, during irradiation, accelerated protons penetrate through the top $MX_2$ basal plane (Fig. 2d top), becoming confined in between two X-M-X layers. Therein, triggered by the catalytic activity of TMDs[14,15,16], the following reaction takes place:

$$2\mathrm{H}^{+} + 2e^{-} \rightarrow \mathrm{H}_2, \quad (1)$$

with electrons $e^{-}$ being supplied from the ground contact (Fig. 2d middle). The subsequent build-up of $H_2$ molecules, stored just beneath the top surface, leads to the local blistering of one X-M-X plane, and eventually to the formation of the domes (Fig. 2d bottom). The above scenario is supported by theoretical studies, which showed that thermal protons remain trapped in the metal plane of X-M-X layers and do not diffuse, thus favoring the accumulation of $H_2$ molecules in the interlayer regions[24]. The formation of localized nucleation sites (*i.e.*, the domes), rather than the establishment of a uniform $H_2$ distribution underneath the top layer, is due to the van der Waals (vdW) forces existing between adjacent planes, which prevent $H_2$ molecules from moving freely. The shape of the domes itself—*i.e.*, their height-to-radius ratio—results from the interplay between: (*i*) the vdW forces that tend to keep the TMD layers together, thus minimizing the dome radius; (*ii*) the stiffness of the top TMD layer, which limits the dome height; and (*iii*) the tendency to expand of the trapped gas, which results in the application of a constant pressure on the dome's walls[6]. We estimate such pressure to be about 10 atm for $R \gtrsim 500$ nm, given the average $T^{\mathrm{H}}_{\mathrm{LV}}$=32.2 K and the $H_2$ phase diagram (see SN1 and Ref. 25). Therefore, the multi-colored look sported by the large domes in Fig. 2a can be ascribed to *Newton's rings* caused by the interference of the light reflected inside the $H_2$-filled spherical volume[26].

## 3. Domes' strain fields

### 3.1 Domes' profile and strain tensor modeling

As noted earlier and shown in Fig. 1f, the average PL peak energy of the domes is about 200 meV lower than that of the corresponding material in the ML form (see also Fig. S5). This is ascribable to the presence of tensile strain, $\varepsilon$, which reaches its maximum $\varepsilon_{\mathrm{m}}$ at the domes' summit. As predicted by Hencky's model[7,27,28],

$$\varepsilon_{\mathrm{m}} = f(\upsilon) \cdot (h_{\mathrm{m}}/R)^2, \quad (2)$$

where $f$ only depends on the material's Poisson ratio, $\nu$, while $R$ and $h_m$ are the dome footprint radius and maximum height, respectively; see Supplementary Note SN2. As reported in Ref. 6, the dome's height-to-radius ratio is largely independent of the dome's dimensions: $h_m$ scales linearly with $R$, and the value of $\varepsilon_m$ in a given material can be estimated by Eq. (2). As shown in Fig. 3a, wherein the experimental (*i.e*, obtained by AFM) values of $h_m$ are plotted as a function of $R$ for six different chemical compositions, the expected linear dependence is verified over a wide span of dome sizes ($R$ between 100 nm and ~3 μm). The average values of $h_m/R$ estimated for each of the investigated materials are: 0.16±0.02 ($MoS_2$), 0.18±0.02 ($MoSe_2$), 0.17±0.02 ($MoTe_2$), 0.16±0.02 ($WS_2$), 0.15±0.01 ($WSe_2$) and 0.13±0.02 ($WTe_2$) yielding the values of $\varepsilon_m$ shown in Fig. 3a and in SN2.

The height profile of the dome and the evolution of the strain tensor across the dome's surface can both be computed via Finite-Element Method (FEM) calculations, performed within the framework of nonlinear membrane theory (see Supplementary Note 3)[27]. As an illustrative case, Fig. 3b (left) shows the comparison between these calculations and the AFM profile of a $MoS_2$ dome with $R$=(3.76±0.12) μm and $h_m$=(618±15) nm, formed after irradiation of a flake with $8\times10^{16}$ protons/cm$^2$.

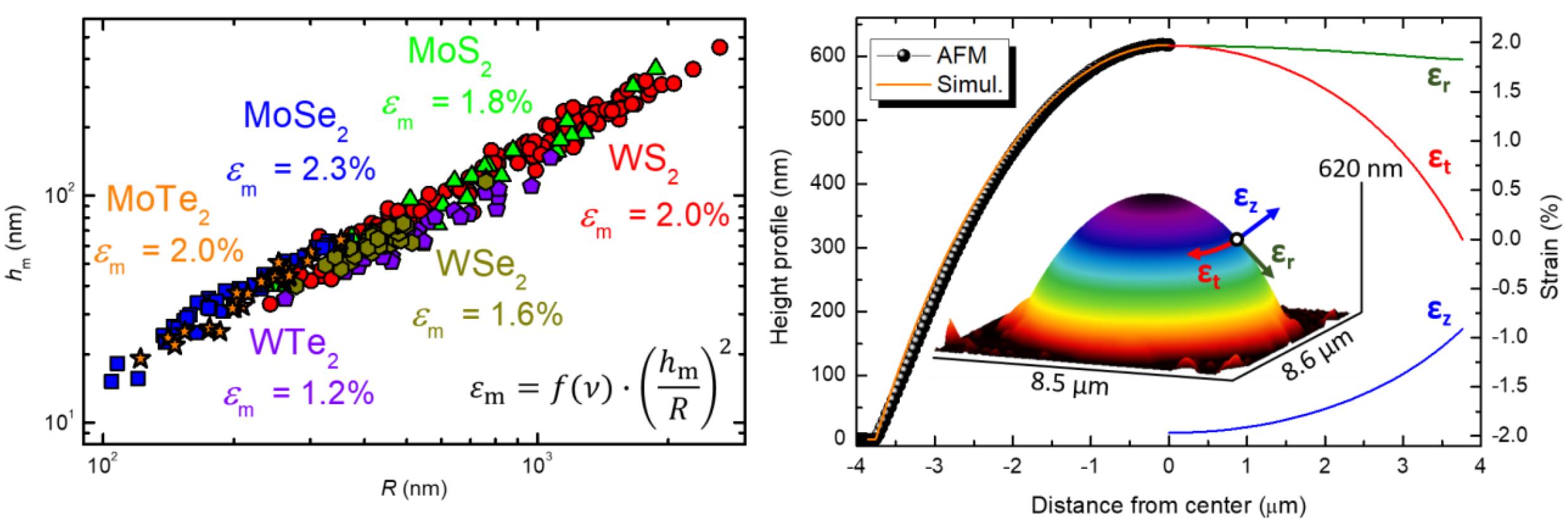

**Figure 3. Dome's shape, strain, and size distribution**. **a,** Maximum height $h_m$ *vs* footprint radius $R$ of $MX_2$ domes. The average $h_m/R$ provides the average dome biaxial tensile strain at the dome apex $\varepsilon_m$, according to the equation reported in the figure. **b,** Left side: Comparison between the height profile of a $MoS_2$ dome measured by AFM (black dots), and the profile obtained by Finite Elements calculations (solid orange line; see Methods). The dome footprint radius and height, as estimated by AFM measurements, are 3.76 μm and 618 nm, respectively. Right side: Components of the strain tensor along the three principal axes of the dome, *i.e.*, in the radial ($\varepsilon_r$, solid olive line), circumferential ($\varepsilon_t$, solid red line), and perpendicular ($\varepsilon_z$, solid blue line) directions. These components are represented as color-coded arrows in the inset, showing a three-dimensional AFM image of the dome.

As summarized in Supplementary Fig. SN3A, FEM simulations correctly reproduce the experimental profile for all the investigated materials. The calculated spatial dependences ($0 \leq r \leq R$) of the principal components of the strain tensor —namely, along the circumferential ($\varepsilon_t$) and radial ($\varepsilon_r$) in-plane directions[27] and along the perpendicular ($\varepsilon_z$) out-of-plane direction —are also displayed in Fig. SN3A and 3b (right): interestingly, strain is isotropic biaxial ($\varepsilon_t = \varepsilon_r = 1.97\%$ for the dome in Fig. 3b) only at the very top of the dome, gradually becoming uniaxial as the circumferential component goes to zero near the dome's edge. In between these two limits, strain is anisotropic ($\varepsilon_t \neq \varepsilon_r$), with a negative perpendicular component accounting for a compression in out-of-plane direction (as detailed in Supplementary Note 3). The complex spatial dependence of the strain tensor across the domes is reflected by the spatial evolution of μ-Raman measurements where, as a general feature, a progressive softening of the in-plane and out-of-plane vibrational modes is observed with increasing strain over the dome. Due to the diffraction-limited size of our laser spot (directly measured and modelled by a Gaussian with standard deviation $\sigma$=0.23±0.01 μm, see Methods), the finer details of this evolution can only be appreciated on large ($R$>2 μm) domes: μ-Raman measurements were therefore performed on $WS_2$, $MoS_2$ and $WSe_2$, where large domes were fabricated, giving results in good agreement with the literature[7] discussed in Supplementary Figs. S6-S8. Our results agree well also with Hencky's model[7,27], yielding $\varepsilon_{\mathrm{m}} = \varepsilon_{\mathrm{t}} = \varepsilon_{\mathrm{r}} = 1.95\%$ for the $MoS_2$ dome shown in Fig. 3b (the same applies to the other $MX_2$ compounds investigated; see SN3). Moreover, the pressure values used in the calculations match rather well, for μm-sized domes, with the ~10 atm estimated from the average $T^{\mathrm{H}}_{\mathrm{LV}}$ obtained in the previous section (see Fig. 2c). For smaller ($R$<<1 μm) domes, the model results in internal pressures in excess of 100 atm—as shown in Figure SN3B and reported in Ref. 6—for which $H_2$ should enter a supercritical fluid phase[29].

## 4. Engineered creation of TMD domes

The detailed knowledge of the evolution of the strain tensor across the domes' surface is particularly relevant since strain is known to alter profoundly the electronic properties -and hence the optical and transport properties- of TMDs, which can lead to, *e.g.*, the emergence of giant pseudo-magnetic fields and to the generation of persistent currents[10,11,12,13]. For this to happen, specific and stable configurations of the strain field need be achieved, and spatial control and durability are therefore necessary. In fact, other methods for the creation of strained TMD bubbles either allow size/position controllability but lack durability (as in Refs. 7, 9) or permit to create durable structures but which lack any spatial ordering (as in Ref. 6), in both cases considerably limiting their potentiality for applications.

As shown in Fig. S9A, indeed, the domes created via proton-irradiation remain intact—and keep exhibiting strong light emission—for temperatures up to 510 K. In addition, the shape and size of these nano/micro-structures remain unchanged with time—in the best cases over more than two years—as illustrated by Figs. S9B-F. This confirms the notable absence of gas-permeable, nm-scale pores in TMD materials, as already reported in the literature[7]; moreover, the exceptional durability of our domes is likely also aided by the strong adhesion forces existing between the ML forming the dome and its parent substrate, as well as by the low permeability to $H_2$ of the latter. In the bulging devices commonly employed to induce strain in TMD membranes[7,8,9] the ML is usually laying directly on a $SiO_2$ substrate. As a result, the devices typically deflate within weeks, mostly due to leaks through the substrate and the TMD/$SiO_2$ interface[7]. Besides durability, the applicative prospects of our $MX_2$ domes are greatly enhanced by the possibility to precisely control their size and position. As illustrated in Fig. 4, indeed, we can engineer the dome formation process, which can be achieved depositing an H-opaque masking layer prior to proton irradiation. The sample displayed in Fig. 4, for example, was patterned by electron-beam lithography with arrays of circular openings with diameter $S$=1, 3, and 5 μm (see Figure S10 and Ref. 30), followed by proton irradiation ($d_H$=4×10$^{16}$ protons/cm$^2$) and by the removal of the H-opaque mask (see Methods). Fig. 4a displays the AFM image of a $WS_2$ array of neatly arranged single domes with average $R$=(0.93±0.07) μm and $h_m$=(0.13±0.02) μm, obtained for $S$=3 μm. The μ-PL signal emitted by the same array is imaged in Fig. 4b (see also the μ-PL map in the inset). The ability to fabricate ordered arrays of light-emitting domes is an inherent advantage of our system with respect to, *e.g.*, the bubbles forming because of the accidental incorporation of contaminant gases between monolayers and their supporting substrates[6,31], which have been shown to be durable[32] but lack any spatial ordering. The dome dimensions can also be engineered by varying the opening size.

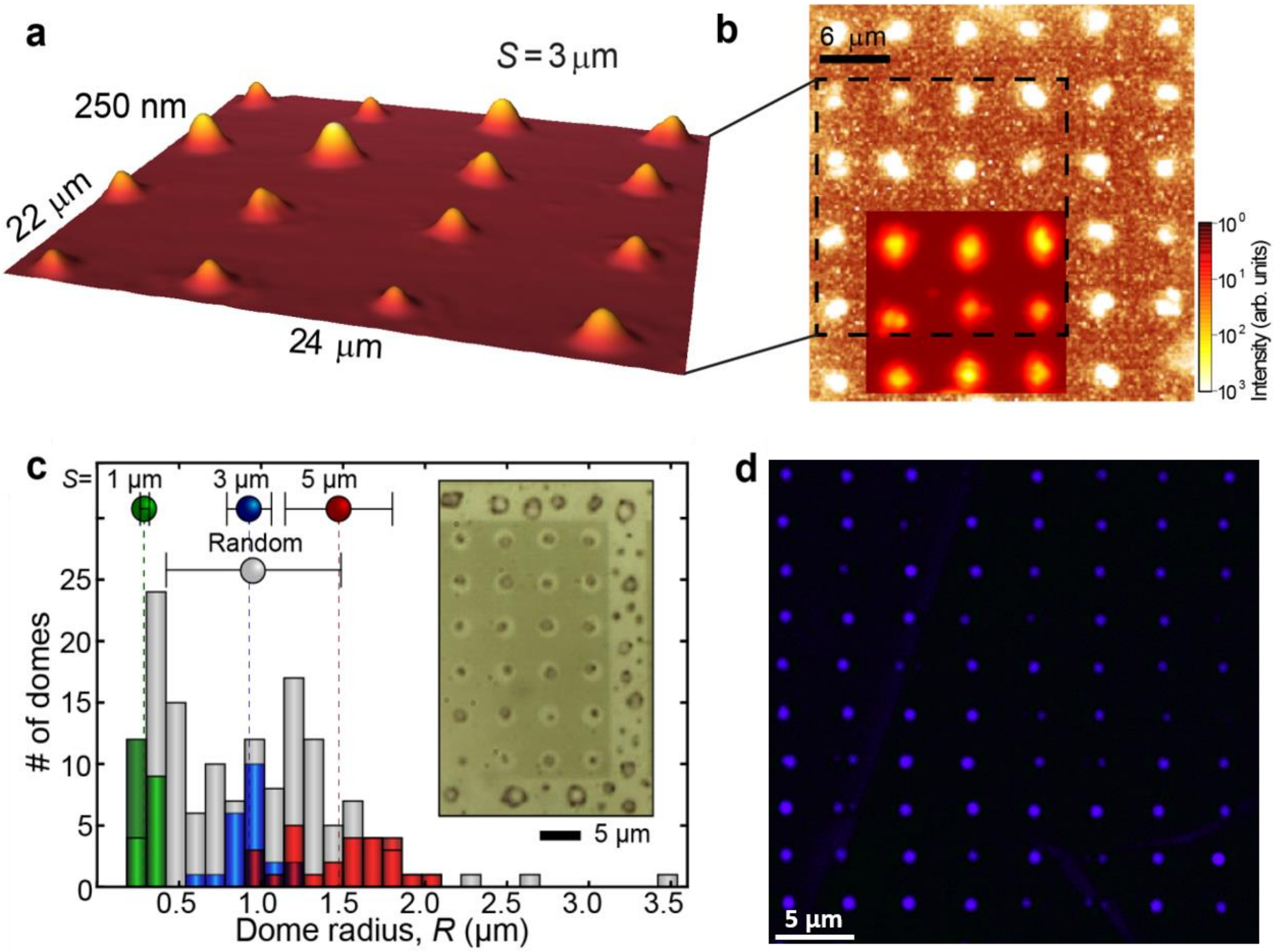

**Figure 4. Controlling the dome size and position**. **a,** AFM image of an array of $WS_2$ domes obtained after $H^+$ irradiation (dose $d_H$=4×10$^{16}$ protons/cm$^2$) of a flake patterned with an H-opaque mask. The mask had openings with diameter *S*=3 μm and was removed before the AFM measure (see also Fig. S10). **b,** RT PL pan-chromatic imaging of the same array in panel (**a**) excited by a 532.2 nm laser. The bottom-left inset is a μ-PL mapping (detection wavelength equal to 689 nm) of a portion of the same array. **c,** Distribution of the footprint radius of domes formed in opening arrays with diameter *S*=1 (green), 3 (blue), and 5 (red) μm. The distribution of randomly formed domes during the same process is given by gray bars. The distribution width is illustrated on top of the figure by horizontal bars. The inset is an optical microscope image comparing random vs ordered (*S*=3 μm) domes. The much narrower distribution of the ordered domes with respect to the random ones can be appreciated. **d,** Second harmonic generation (SHG) map collected on an array of ordered $WS_2$ domes obtained after proton irradiation (dose $d_H$=1.5×10$^{17}$ protons/cm$^2$) of a flake patterned with an H-opaque mask. The mask had elliptical openings with 2- and 1-μm semi axes and 4-μm center-to-center distance.

In fact, Fig. 4c shows the distributions of the dome footprint radii grouped into randomly formed and ordered dome subsets obtained during the *same* proton irradiation process. It is worth noting that the domes formed using the lithographic approach sport a remarkably narrower size distribution, where the size is clearly determined by the diameter of the aperture. Such distribution gets even narrower with decreasing the dome size. Finally, in the ordered arrays, the average dome volume scales with the surface area ($0.25\pi S^2$) available to the reaction (1) (see Fig. S11), thus supporting the hypothesis of the dome formation as a catalysis-driven process. We point out that controlling the dome position and size over large areas is valuable for the creation of efficient site-controlled photon emitters, as well as in many other situations. For example, TMD MLs have been shown to give rise to efficient second-harmonic generation (SHG), due to their broken inversion symmetry[33,34]. This property is retained by our domes -as demonstrated in Fig. 4d for an ordered array of $WS_2$ domes excited at 900 nm (see Methods)- which could thus conceivably act as site-controlled SHG "hot spots", ideal for the integration with specifically designed photonic crystal cavities[35,36]. One should also notice that the intense second harmonic signal featured by the domes in panel d indicates a $> 96\%$ formation yield of one-ML-thick domes.

## 5. Conclusions and perspectives

ML TMDs have been demonstrated to possess intriguing optoelectronic and mechanical properties, which renders them suitable for flexible and innovative optoelectronic devices.

Furthermore, they have emerged as a fascinating class of materials for catalysis[15]. Here, we demonstrated that the catalytic properties of TMDs can be activated by simply irradiating bulk flakes with a low energy proton flux, which leads to the formation and accumulation of hydrogen and allows to control the electronic properties of ML TMDs at the nano- and micro- scale through the creation of highly strained domes. These structures can be formed at the desired location and with controllable size on six different compounds. Since we can precisely control the amount of $H_2$ in single domes, they can act as microscopic calibration gauges for $H_2$ sensors or for controlled reactive gas delivery (see Fig. S12) in nano-reactors[37] with unprecedented accuracies. The engineered formation of domes also allows the creation of arrays of linear and non-linear light emitters and of periodic strain-fields, as here demonstrated. Finally, it should be noted that several other 2D crystals—such as graphene, h-BN, and phosphorene—have been found to be virtually transparent to protons, according to both experiments[38] and theoretical calculations[24]. Therefore, their deposition on top of bulk TMD samples before proton irradiation should not hinder the dome formation process, and the controlled formation of (ordered) TMD domes can thus be exploited to define a template for modulating strain in a much wider range of two-dimensional materials, leading, *e.g.*, to the emergence of giant pseudo-magnetic fields in graphene, and creating the conditions for the generation of dissipation-less electrical currents[10,11,12,13].

**Movie 1**

Optical microscope movie of a $WS_2$ sample irradiated with an $H^+$ dose $d_H = 8\times10^{16}$ protons/cm$^2$. The movie (acquired by employing a 50× objective with NA=0.5) was taken as the temperature was raised from $T = 19.75$ K to $T = 45.05$ K. The temperature frame step is 10 mK. The counter visible on the top-right corner of the movie displays the sample temperature. At $T$~28 K the domes start inflating due to the liquid-vapor transition of $H_2$.

## Methods

### 1 Sample preparation and proton irradiation procedure

$MX_2$ (M=W, Mo; X=S, Se, Te) bulk flakes are mechanically exfoliated by the scotch tape method on various Si substrates: bare, $SiO_2$- and Au-covered. The samples are then mounted in a vacuum chamber secured on a holder in such a way to guarantee the thermal and electrical (ground) contact between the holder and the sample. The chamber is evacuated at $p = 1.0\times10^{-6}$ mbar and the temperature of the sample holder is set at 150° C. Meanwhile, protons are produced in an ionization chamber and accelerated by a system of grids that generates an $H^+$ beam steered onto the sample surface with energies between about 5 and 500 eV (in the present study an energy < 100 eV was employed). The total impinging $H^+$ dose is determined by setting the exposure time once the proton flux is fixed. Once the target dose is obtained, the sample holder is slowly cooled down to room temperature and the chamber is brought to atmospheric pressure. The duration of the whole process is of the order of few hours. A sketch of the hydrogenation setup is shown in Fig. S13A.

### 2 Liquid-to-vapor phase transition measurements

The measurements were performed by mounting the sample in a closed-cycle cryostat by Montana Instruments allowing the use of a 50× objective with NA=0.5 mounted externally to the cryostat optical window. The image of the sample was monitored on a CCD camera to record in real time the sample image with a rate of 10 frames per second. The sample was mounted on a special copper holder featuring a thermal link to the cryostat platform (the coldest point of the system) thus ensuring optimal and fast heat exchange with the sample. The temperature was varied from 5 K to more than 50 K using a calibrated heater. The sample temperature *T* was measured by a calibrated 4-point thermometer mounted as close as possible to the samples and *T* was increased at a low rate equal to ~40 mK/s. The liquid-vapor phase transition temperature was established by subsequent careful visual inspection of the videos acquired during the *T* sweep. More details on the isotopic shift measurements are given in SN1.

**3 Optical measurements**

Photoluminescence (PL) and Raman experiments were performed using the following equipment. The excitation laser was provided by two different single-frequency Nd:$YVO_4$ lasers (V8 Verdi by Coherent Inc. and DPSS series by Lasos) with emission wavelength equal to 532.2 nm. The PL emission was spectrally analyzed by means of a 750-mm focal length monochromator ACTON SP750i equipped with 3 gratings (300, 600, and 1200 groove/mm) and detected by a back-illuminated Si CCD Camera (model 100BRX) by Princeton Instruments. The μ-Raman spectral resolution was 0.7 $cm^{-1}$ and the μ-PL spectral resolution was 0.2 meV. The laser light was filtered out by a very sharp high-pass Razor edge filter at 535 nm (Semrock).

For micro(μ)-PL/Raman on single domes or ML, a confocal microscope was used to excite and collect the light. Several microscope objectives (100× NA=0.9, 50× NA=0.5, 20× NA=0.4) were employed depending on the specific experiment.

For μ-Raman mapping (Figs. S6-S8), we used an Olympus 100× objective with NA=0.9. The minimum scanning step was 80 nm using a compact open loop piezoelectric xy-scanner (Attocube). The mapping was performed along a dome diameter to reduce the data acquisition time and thus increase the accuracy of the measurements. We checked that similar results were obtained also along different diameters of the domes.

The laser spot size resulting from the NA=0.9 objective employed for μ-Raman mapping was experimentally determined as follows: the laser was scanned across a reference sample, lithographically patterned with features of known width (1 μm). The intensity of the reflected light was fitted with the ideal reflectance profile, convolved with a Gaussian peak. The standard deviation of this peak, obtained as a fitting parameter, provides our estimate of $\sigma$=0.23±0.01 μm.

For PL measurements on dome ensembles, the excitation laser was focalized by a lens of 20 cm focal length positioned laterally with respect to the luminescence-collecting objective. In this manner, the samples were excited over an area in excess of 100×100 $\mu m^2$. The PL emission was collected by a 10× Leitz objective with NA=0.2, yielding a PL signal averaged over a large part of the sample under study and allowing us to get statistically reliable. This applies to Fig. 1f (dashed line) and Fig. S9A. The same set-up was used for μ-PL imaging (Fig. 1a and 4b) by collecting the luminescence through a low-aperture microscope objective and monitoring the pan-chromatic PL intensity with a CCD.

Low-temperature μ-PL mapping and imaging experiments were performed using a Montana cooler cryostation able to reach the lowest temperature of 4 K.

The circular polarization-resolved μ-PL experiments (Fig. S3) were performed by using a He-Ne laser by UniPhase as excitation source ($\lambda_{exc}$=632.8 nm) using the setup and procedure described in Fig. S13B.

All optical measurements were intensity-normalized by the set-up spectral response by using a calibrated black-body source at 3050 K.

The second harmonic generation (SHG) measurements (Fig. 4d) were conducted with a confocal optical microscope (Carl-Zeiss LSM 780) equipped with a 50× optical object (Epiplan-Neofluar 0.85, DIC). A Ti:Sa laser from Spectra-Physics (100 fs, 80 MHz, wavelength tuneable between 690 and 1040 nm) was used to pump the sample. The laser wavelength was set at 900 nm.

**4 Atomic force microscope measurements**

AFM measurements were performed in tapping mode, by using two different AFM instruments that provided the same results within the experimental uncertainty. In one case, we used a Dimension Icon microscope with a Nanoscope V controller (Bruker), equipped with R-Tespa tips. In another case, we used a Veeco Digital Instruments Dimension D3100 microscope equipped with a Nanoscope IIIa controller, employing Tapping Mode monolithic silicon probes with a nominal tip curvature radius of 5-10 nm and a force constant of 40 N/m.

All the data were analyzed with the Gwyddion software and the convolution of the tip used was duly taken into account. All the data were taken at room temperature and under the same ambient conditions (atmospheric pressure).

**5 Electron-beam lithography patterning**

The fabrication of H-opaque masks was performed by means of electron-beam lithography employing a Vistec EPBG 5HR system working at 100 kV. A hydrogen silesquioxane (HSQ) negative-tone *e*-beam resist was employed because of its property to be H-opaque under the irradiation conditions used in this work. Ordered arrays of octagonal openings with the desired diameter were patterned on an 80-nm thick HSQ masking layer deposited on top of the sample surface. An electron dose of 300 μC/cm$^2$ and an aqueous development solution of tetramethyl ammonium hydroxide at 2.4% were used for the patterning of the HSQ masks. After the H treatment, the HSQ masks were removed by chemical etching in an aqueous solution of potassium borates at 5-15% and potassium hydroxide at 2%, which dissolves HSQ but does not attack the transition metal dichalcogenide samples (see Figs. 4 and S10).

**Author contributions**

D.T., E.B., M.F. and A.P. fabricated the domes, performed and designed the optical experiments, and conducted the analysis regarding the optical and microscopy data. E.B. performed and analyzed the isotopic shift measurements. G.P. designed and performed the sample lithographic processing and the microscopy measurements. T.Y. and Y.L. made the mechanical modeling calculations. E.P. and S.S. contributed at the initial stages of this work. B.L., C.Z., Y.Z. prepared the samples and performed part of the optical experiments. Y.L. coordinated the work at the Australian National University. A.P. and M.F. wrote the manuscript. A.P. conceived the research and coordinated the work at Sapienza Università di Roma.
All authors discussed the results and commented on the manuscript.

**List of Supplementary Material**

Supplementary Figures S1-S13 and Supplementary Note SN1, SN2, SN3.

Movie 1.